\documentclass{nature_withfigs}

\usepackage{amsmath,amsfonts,amssymb}
\usepackage{graphicx}

\title{Designer atoms for quantum metrology}

\author{C.~F.~Roos$^{1,2}$, M.~Chwalla$^{1}$, K.~Kim$^{1}$, M.~Riebe$^{1}$, and R.~Blatt$^{1,2}$}

\begin{document}

\maketitle

\begin{affiliations}
\item Institut f{\"u}r Experimentalphysik, Universit{\"a}t
Innsbruck, Technikerstr.~25, A-6020 Innsbruck, Austria \item
Institut f{\"u}r Quantenoptik und Quanteninformation,
{\"O}sterreichische Akademie der Wissenschaften,
Otto-Hittmair-Platz~1, A-6020 Innsbruck, Austria
\end{affiliations}

\begin{abstract}
Entanglement has been recognized as a key resource for quantum
computation\cite{Nielsen00} and quantum
cryptography\cite{Gisin02}. For quantum metrology, use of
entangled states has been mainly discussed as a means of improving
the signal-to-noise
ratio\cite{Giovannetti04,Giovannetti06,Bollinger96} and first
demonstration experiments were performed\cite{Leibfried04}. In
addition, entangled states were used in experiments for efficient
quantum state detection\cite{Schmidt05} and for the measurement of
scattering lengths\cite{Widera04}. In quantum information
processing, manipulation of individual quantum bits allows for the
tailored design of specific states that are insensitive against
detrimental influences of an environment\cite{Kielpinski01}. Such
decoherence-free subspaces\cite{Lidar98} protect quantum
information and yield significantly enhanced coherence
times\cite{Roos04}. Here, we demonstrate precision spectroscopy
using a decoherence-free subspace with specifically designed
entangled states\cite{Roos05} to obtain the electric quadrupole
moment of Ca${^+}$ to be used for frequency standard applications.
We find that entangled states are not only useful for enhancing
the signal-to-noise ratio in frequency measurements but that
moreover a suitably designed pair of atoms will allow clock
measurements in the presence of strong technical noise. The
applied technique makes explicit use of non-locality as an
entanglement property and constitutes thus a new paradigm for
designed quantum metrology.
\end{abstract}

Metrology seeks to relate all measurements to a precision
determination of frequencies. The most precise measurements today
are obtained from measuring atomic transition frequencies by
monitoring the phase evolution of a superposition of the
pertaining states. This is achieved by applying Ramsey's
interferometric technique\cite{Ramsey50} that can be
generalized\cite{Bollinger96} to (maximally) entangled states:
Under free precession, the two-atom state
$\Psi=\frac{1}{\sqrt{2}}(|u_1\rangle|u_2\rangle+|v_1\rangle|v_2\rangle)$
evolves into the state
$\Psi(\tau)=\frac{1}{\sqrt{2}}(|u_1\rangle|u_2\rangle+\exp(i\lambda_\phi\tau)|v_1\rangle|v_2\rangle)$
where the phase evolution rate
$\lambda_\phi=[(E_{u_1}+E_{u_2})-(E_{v_1}+E_{v_2})]/\hbar$ is
proportional to the differences in atomic energies $E_{u_k}$,
$E_{v_k}$ of the involved levels. The real part of the phase
factor $e^{i\lambda_\phi \tau}$ is measured by projecting the
atoms onto the states
$|\pm\rangle_k=\frac{1}{\sqrt{2}}(|u_k\rangle\pm|v_k\rangle)$ and
measuring the parity operator
$P=\hat{P}_{(++)}+\hat{P}_{(--)}-\hat{P}_{(+-)}- \hat{P}_{(-+)}$,
where $\hat{P}_{(\pm,\pm)}$ denotes the projector onto the state
$|\pm\rangle_{1}\otimes|\pm\rangle_{2}$. If $\Psi$ belongs to a
decoherence-free subspace, free precession times $\tau$ of several
seconds\cite{Haeffner05,Langer05} allow for highly accurate phase
estimation.

In atomic optical frequency standards based on single
hydrogen-like trapped ions\cite{Madej01}, the transition frequency
from the $S$ ground state to a metastable $D_j$ state is measured
by an optical frequency comb\cite{Diddams04}. The $D_j$ state's
atomic electric quadrupole moment interacting with residual
electric quadrupole fields\cite{Itano00} gives rise to frequency
shifts of a few Hertz. The shift of the Zeeman sublevel
$|D_j,m_j\rangle$ in a quadrupole field $\Phi(x,y,z)=
A(x^2+y^2-2z^2)$ is given by
\begin{equation}
\label{shiftformula}
\hbar\Delta\nu=\frac{1}{4}\frac{dE_z}{dz}\Theta(D,j)\frac{j(j+1)-3{m_j}^2}{j(2j-1)}
(3\cos^2\beta-1),
\end{equation}
where $dE_z/dz=4A$ is the electric field gradient along the
potential's symmetry axis $z$, $\beta$ denotes the angle between
the quantization axis $z$ and $\Theta(D,j)$ expresses the strength
of the quadrupole moment in terms of a reduced matrix
element\cite{Itano00}.

\begin{figure}
\includegraphics[width=1\textwidth]{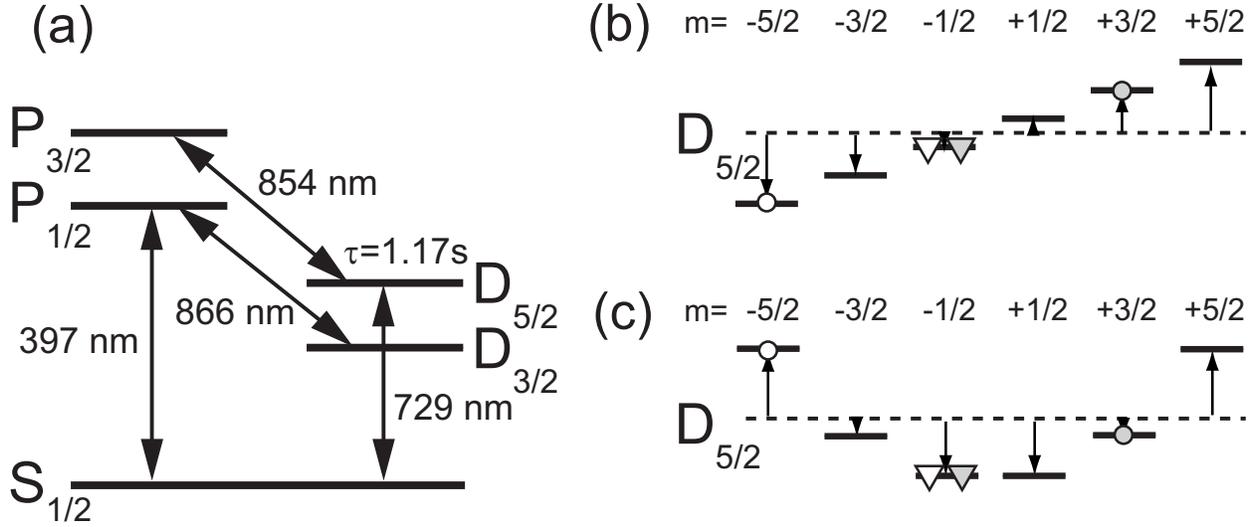}
\caption{\label{levels} Relevant atomic levels of the
$^{40}$Ca$^+$.  (a) Bell states are prepared by coherently
exciting the $S_{1/2}\leftrightarrow D_{5/2}$ quadrupole
transition. The $S_{1/2}\leftrightarrow P_{1/2}$ is used for
quantum state read-out, lasers at 866 and 854 serve as repumping
and quenching lasers. (b) Energy level shift of the
$|m\rangle\equiv |D_{5/2},m\rangle$ states in a weak magnetic
field. (c) Energy level shift caused by an electric quadrupole
potential. Energy levels occurring in the state
$\Psi_1=\frac{1}{\sqrt{2}}
(|\circ\rangle|\bullet\rangle+|\triangledown\rangle|\blacktriangledown\rangle)$
(eq.\,(\ref{state})) are denoted by the symbols
$\circ,\triangledown,\bullet,\blacktriangledown$, the open
(filled) symbols corresponding to levels occupied by atom 1 (2),
respectively. The combined Zeeman energy of the
$|\circ\rangle|\bullet\rangle$-states equals the energy of the
$|\triangledown\rangle|\blacktriangledown\rangle$-states, making
$\Psi_1$ immune against magnetic field noise. Since an electric
quadrupole field shifts the $|\circ\rangle|\bullet\rangle$- and
$|\triangledown\rangle|\blacktriangledown\rangle$-states in
opposite directions, $\Psi_1$ is suitable for a measurement of the
quadrupole shift.}
\end{figure}

Recently, quadrupole moments have been measured for $^{88}$Sr$^+$,
$^{199}$Hg$^+$ and $^{171}$Yb$^+$ with a precision ranging from
about 4$\%$ to 12$\%$ (refs.
\cite{Barwood04,Oskay05,Schneider05}). In these single-ion
experiments, narrow-linewidth lasers are employed to detect
electric quadrupole shifts by measuring the transition frequency
from the electronic ground state to the metastable state.
Alternatively, the quadrupole shift could also be measured by
performing Ramsey experiments with an ion prepared in a
superposition of Zeeman $D_j$ sublevels. Both measurement schemes
are subject to phase decoherence. Laser frequency noise limits the
coherence time if the $S$ state is used as the reference state.
Magnetic field noise giving rise to first-order Zeeman shifts
renders the measurement scheme based on $D$ state superpositions
difficult. However, both sources of phase noise can be eliminated
by replacing the single ion by a two-ion entangled state prepared
in a decoherence-free subspace. Ramsey experiments with the
two-ion Bell state
$\Psi=\frac{1}{\sqrt{2}}(|m_1\rangle|m_2\rangle+|m_3\rangle|m_4\rangle)$,
where the magnetic quantum numbers $m_i$ of the states
$|m_i\rangle\equiv|D_{j},m_i\rangle$ satisfy $m_1+m_2=m_3+m_4$,
are not affected by fluctuations of the magnetic fields to first
order since both parts of the superposition are Zeeman-shifted by
the same amount (see Fig.\,\ref{levels}b). Also, frequency noise
of the laser used for preparing $\Psi$ is relevant only during the
comparatively short state preparation and read-out steps but not
during the long interrogation period. For suitably chosen values
of $m_1,\ldots m_4$, a decoherence-immune state can be designed
that is sensitive to the electric quadrupole shift
(Fig.\,\ref{levels}c).

The isotope $^{40}$Ca$^+$ has a single valence electron and no
hyperfine structure (Fig.\,\ref{levels}a). For its metastable
$3d\,^2D_{5/2}$ state (lifetime $\tau_{D_{5/2}}=1.168(7)$~s), a
quadrupole moment of $1.917\,\mbox{ea}_0^2$ was
calculated\cite{Itano06,Sur06}. For the experiment, two
$^{40}$Ca$^+$ ions are trapped along the axis of a linear ion
trap. By applying a static (dc) voltage ranging from 500 to 2000
$V$ to the tip electrodes\cite{SchmidtKaler03}, axial
center-of-mass (COM) mode trap frequencies $\omega_z$ ranging from
850 to 1700~kHz are achieved, leading to distances between the
ions from $6.2\,\mu m$ to $3.9\,\mu m$. A CCD camera images the
ions' fluorescence. The degeneracy of the Zeeman states is lifted
by applying a magnetic bias field of 2.9\,G. The ions are cooled
to the ground state of the axial COM mode by Doppler and sideband
cooling\cite{Roos99}. For coherent quantum state manipulation, the
ions are individually excited on the $|S\rangle\equiv
S_{1/2}(m=-1/2)\leftrightarrow |D\rangle\equiv D_{5/2}(m=-1/2)$
transition by a tightly focussed laser beam (laser linewidth
$\approx 150\,$Hz) resonant with either the carrier transition or
the upper motional sideband. For a detailed description of the
experimental setup, see ref.\cite{SchmidtKaler03}.

The Bell state
$\Psi^\pm_{SD}=(|S\rangle|D\rangle\pm|D\rangle|S\rangle)/\sqrt{2}$
is created with a fidelity of about $90\%$ by a sequence of three
laser pulses\cite{Roos04}. Additional carrier $\pi$~pulses
transfer the entanglement into the $D_{5/2}$ Zeeman state
manifold, thus generating the state
\begin{equation}
\label{state}
\Psi_1=\frac{1}{\sqrt{2}}(|-5/2\rangle|+3/2\rangle+|-1/2\rangle|-1/2\rangle).
\end{equation}
This magnetic field-insensitive state is used for a measurement of
the quadrupole shift. In the vicinity of the trap center, the dc
voltage applied to the tip electrodes creates a rotationally
symmetric electric quadrupole potential which shifts the energy of
the constituents of the superposition state $\Psi_1$ by
$\hbar\Delta_1=24/5\,\hbar\delta$ with respect to each other.
Here, $\hbar\delta$ is the shift that a single ion in state
$|-5/2\rangle$ would experience. The two-ion shift is bigger by
$12/5$ because two ions contribute that are prepared in substates
which shift in opposite directions. In addition, the presence of a
second ion doubles the electric field gradient at the location of
the other ion\cite{Roos05}. For the shift measurement, $\Psi_1$ is
let to evolve into
$\Psi_1(\tau)=\frac{1}{\sqrt{2}}(|-5/2\rangle|+3/2\rangle+\exp(i\Delta_1\tau)|-1/2\rangle|-1/2\rangle)$
and $\cos(\Delta_1\tau)$ is measured for $\tau$ ranging from 0 to
300~ms.
\begin{figure}
\includegraphics[width=1\textwidth]{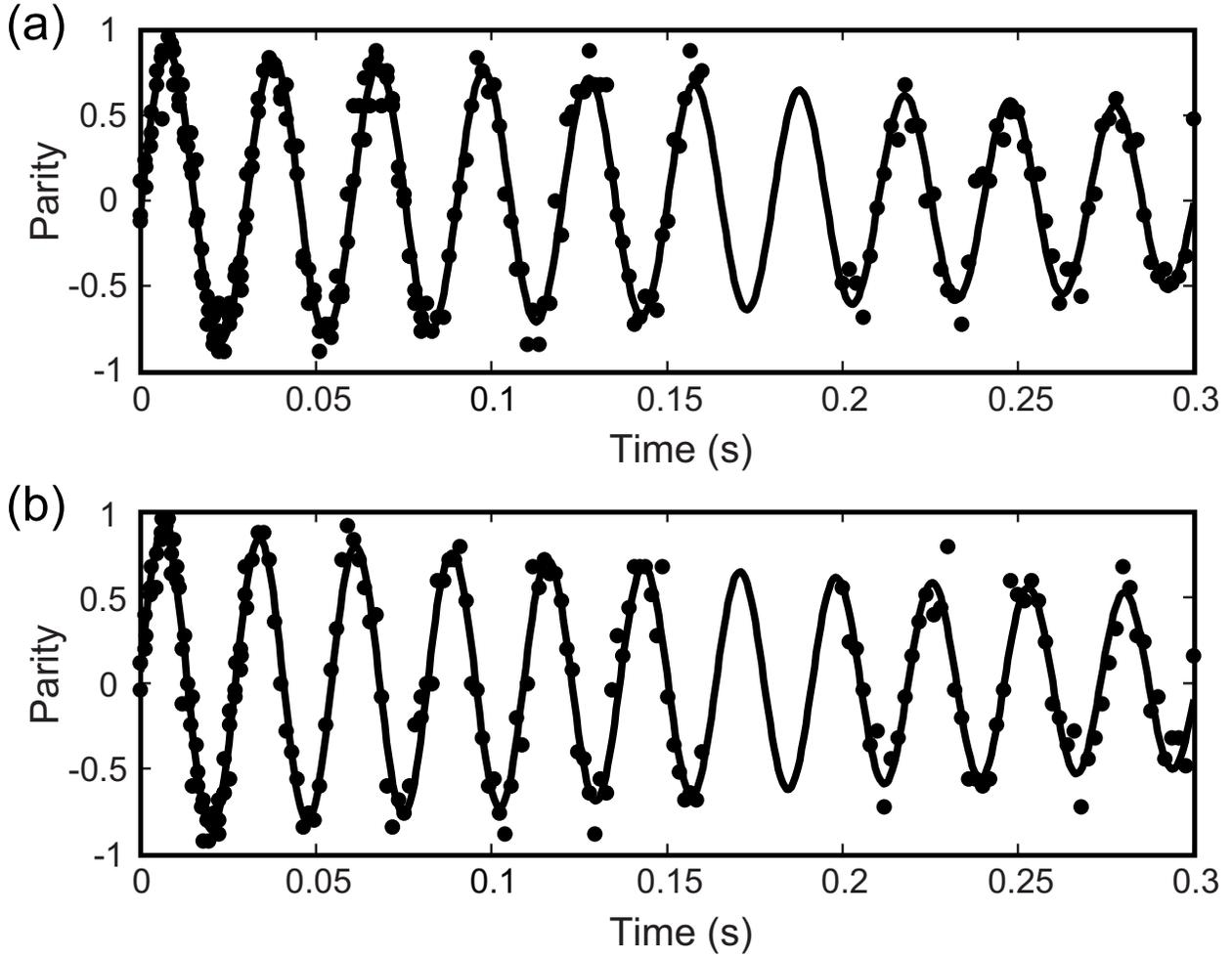}
\caption{\label{parityoscillations} Parity oscillations of the
entangled states $\Psi_1$ and $\Psi_2$ at $U_{tips}=750$~V tip
voltage. The experiment is repeated 100 times for each data point.
An exponentially damped sinusoidal function is fitted to the data.
(a) For $\Psi_1$, the oscillation frequency is
$\Delta_1=(2\pi)\,33.35(3)$~Hz and the damping time $587(70)$~ms.
(b) For $\Psi_2$, the fit yields $\Delta_2=(2\pi)\,36.52(4)$~Hz
and a decay time $525(60)$~ms. For each plot, five different
datasets were merged. No data were taken covering waiting times
around 170 ms. To exclude slow drifts of the initial parity over
time, data for waiting times up to 20 ms were repeatedly taken.}
\end{figure}
Figure \ref{parityoscillations}a shows the resulting parity
oscillations\cite{Roos04} at a frequency
$\Delta_1=(2\pi)\,33.35(3)$~Hz. Disentanglement of $\Psi_1(\tau)$
by spontaneous decay manifests itself as damping of the
oscillations with a damping time constant $\tau_d=587(70)$~ms
close to the expected value $\frac{1}{2}\tau_{D_{5/2}}=584$~ms.
Parity oscillations of the state
$\Psi_2(\tau)=\frac{1}{\sqrt{2}}(|3/2\rangle|-5/2\rangle+\exp(i\Delta_2\tau)|-1/2\rangle|-1/2\rangle)$
are shown in Fig.\,\ref{parityoscillations}b. Its oscillation
frequency $\Delta_2=(2\pi)\,36.52(4)$~Hz slightly differs from
$\Delta_1$ because of a magnetic field gradient in the direction
of the ion crystal that gives rise to an additional contribution
to the parity signal\cite{Roos05}. Both phase oscillation
contributions can be separately determined by taking the average
$\Delta=(\Delta_1+\Delta_2)/2$ and the difference
$\Delta_{B^\prime}=|\Delta_1-\Delta_2|/2$ of the signals. By
recording parity oscillations at different tip voltages
$U_{tips}$, we measure the quadrupole shift $\Delta/(2\pi)$ as
function of the electric field gradient and observe a linear
change in $\Delta(U_{tips})$.
\begin{figure}
\includegraphics[width=1\textwidth]{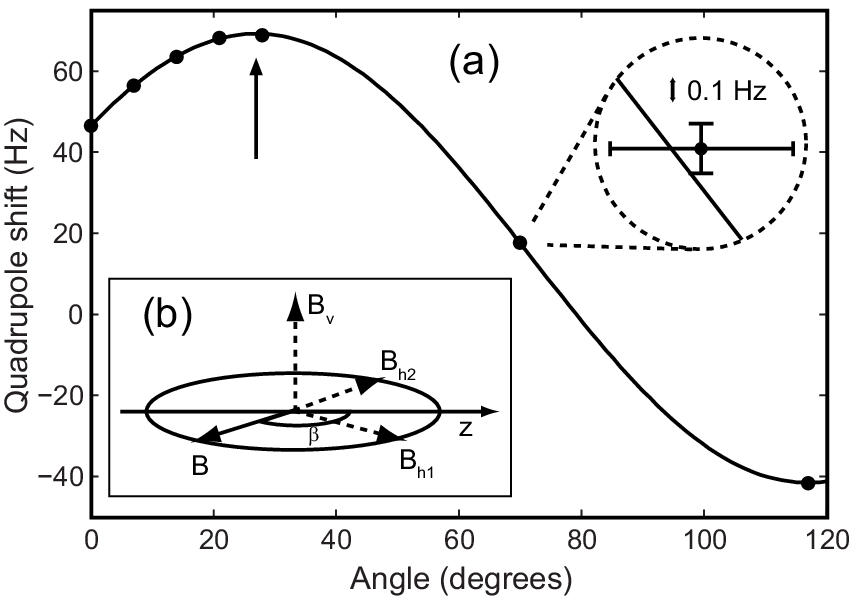}
\caption{\label{angle} Angular dependence of the quadrupole shift.
(a) Quadrupole shift $\Delta/(2\pi)$ at $U_{tips}=1000$~V as a
function of the magnetic field orientation $\beta-\beta_0$. The
data are fitted by
$\Delta=\Delta_a+\Delta_b\cos^2(\beta-\beta_0)$, yielding
$\beta_0=26.9^\circ$. The arrow indicates the angle corresponding
to $\beta=0$. For the determination of the quadrupole moment,
shift measurements are performed at $\beta=0$ since errors in
$\beta$ enter the calculation only in second order. Note that the
error bars are smaller than the point size. The inset in the upper
right corner shows an enlarged data point.  (b) Geometry of
magnetic field coils. The angle $\beta$ is varied by appropriately
changing the current in pairs of coils producing the magnetic
fields $B_{h1},B_{h2}$. Field $B_v$ is used to null residual
fields in the perpendicular direction.}
\end{figure}

For a precise determination of the quadrupole moment, the
dependence of the quadrupole shift on the orientation of the
magnetic field should be minimized. We investigate the angle
dependence of $\Delta$ by varying the orientation of the magnetic
field in a plane that also contains the z-axis of the trap.
Initialization of the ions in a pure state prior to coherent
manipulation is achieved by optical pumping on the quadrupole
transition (see Methods). Generally, it cannot be excluded that
the quadrupole field has imperfect rotational symmetry. In this
case, the factor $3\cos^2(\beta)-1$ in eq.(\ref{shiftformula})
needs to be replaced by
$(3\cos^2\beta-1)-\epsilon\sin^2\beta\cos(2\alpha)$ where
$\epsilon$ characterizes the asymmetry and $\alpha$ its
direction\cite{Itano00}. Note, however, that the additional term
vanishes for $\beta=0$. Figure \ref{angle} shows the sinusoidal
variation of $\Delta$ measured at $U_{tips}=1000$~V as a function
of the angle. At the angle that maximizes $\Delta$, another
maximization is performed in the perpendicular direction to
determine the direction corresponding to $\beta=0$.
\begin{figure}
\includegraphics[width=1\textwidth]{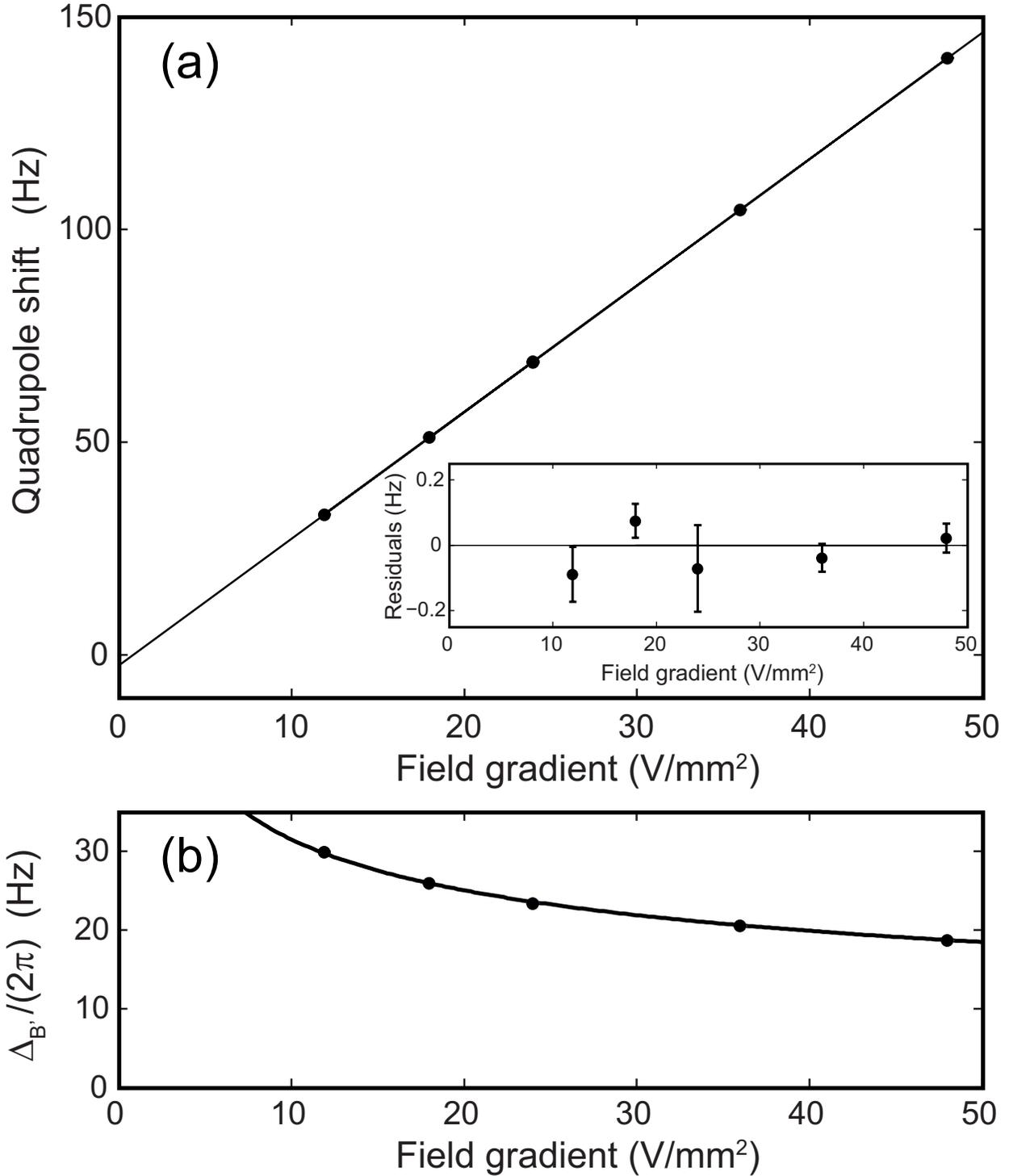}
\caption{\label{quadshift} Level shifts as a function of the
applied external electric field gradient $dE_z/dz$. (a) Quadrupole
shift. The offset at $dE_z/dz=0$ is mainly due to the second-order
Zeeman effect. The quadrupole moment is proportional to the slope
which is measured with an uncertainty of less than $0.1\%$. The
inset shows the deviation of the data points from the linear fit.
(b) Phase oscillation rate $\Delta_{B^\prime}/(2\pi)$ due to the
magnetic field gradient calculated from $|\Delta_2-\Delta_1|/2$.
Since the distance $d$ between the ions decreases at higher
electric field gradients, $\Delta_{B^\prime}\propto d\propto
(dE_z/dz)^{-1/3}$.}
\end{figure}
Figure \ref{quadshift}a displays the quadrupole shift measured at
$\beta=0$ as a function of the electric field gradient $dE_z/dz$.
Using $dE_z/dz=-m\omega_z^2/e$, the gradient is calibrated by a
measurement of the ion's oscillation frequency $\omega_z$. By
fitting a straight line
$\Delta/(2\pi)=\Delta_0/(2\pi)+a\frac{dE_z}{dz}$ to the data, we
determine the slope $a=2.975(2)$~Hz\,mm$^2$/V and the offset
$\Delta_0/(2\pi)=-2.4(1)$~Hz. The major part of the phase
oscillation frequency $\Delta_0$ at $\frac{dE_z}{dz}=0$ is caused
by the second-order Zeeman effect that contributes
$\Delta_{B^2}/(2\pi)=-2.9$~Hz at the bias field of 2.9~G. The
remaining part $\Delta_0-\Delta_{B^2}$ is attributed to a residual
quadrupole field caused by stray charges, ac-Stark shifts are
insignificant. The quadrupole moment
$\Theta(3d,5/2)=\frac{5}{12}ha$ is proportional to the slope $a$.
While the uncertainty in the value of $a$ is smaller than
$10^{-3}$, the main uncertainty in $\Theta(3d,5/2)$ is due to
errors in the determination of the angle $\beta$. Assuming that
the angle can be determined with an accuracy of 3~degrees (see
Methods), we determine the quadrupole moment to be
$\Theta(3d,5/2)=1.83(1)$ea$_0^2$.

These results show the viability of the approach of designing
specific entangled states for precision measurements. For
frequency standard applications, minimization of the quadrupole
shift is imperative\cite{Margolis04,Dube05}. Note that, by using
designed entangled states, the quadrupole shift could be cancelled
by averaging the results over states with different magnetic
quantum numbers\cite{Roos05}. The design approach demonstrated
here exhibits a new paradigm for metrology: while previously the
use of entangled states was to enhance the signal-to-noise ratio,
the use of a designed pair of entangled ions allows an atomic
clock measurement to be performed in a subspace free of magnetic
field noise. This is achieved with unprecedented precision on an
even isotope which opens the perspective for optical frequency
standards based on a whole variety of easily accessible atomic
systems not considered so far.

\begin{methods}
\subsection{Frequency-resolved optical pumping.} Due to the
magnetic bias field, the ions can be prepared in the
$|S_{1/2},m=-1/2\rangle$ state for an arbitrary orientation of the
magnetic field by frequency-resolved optical pumping on the
quadrupole transition. For this, the
$|S_{1/2},m=+1/2\rangle\leftrightarrow |D_{5/2},m=-3/2\rangle$
transition is excited while the repumping lasers on the
$D_{5/2}\leftrightarrow P_{3/2}$ and $D_{3/2}\leftrightarrow
P_{1/2}$ transitions are switched on at the same time. After an
excitation of 1~ms duration, at least $99.9\%$ of the population
has been pumped to the $|S_{1/2},m=-1/2\rangle$ state.
\subsection{Orientation of the magnetic field.} The magnetic
field $\vec{B}$ is set by passing currents $I_{h_1},I_{h_2},I_v$
through three mutually orthogonal pairs of coils (see Fig.
\ref{angle} b). We calibrate the field coils and null the offset
fields by measuring the Zeeman shift of the
$|S_{1/2},m=-1/2\rangle\leftrightarrow|D_{5/2},m=-1/2\rangle$
transition as a function of the currents. We estimate that the
offset fields can be zeroed to a value below 1\% of the applied
bias field. After nulling the field in the v-direction, we map out
the values of $(I_{h_1},I_{h_2})$ that yield a constant Zeeman
shift and fit an ellipse to the data. These current values rotate
$\vec{B}$ in a horizontal plane that also contains the symmetry
axis (z) of the ion trap. A similar procedure is applied for
calibrating the v-coil.

\subsection{Alignment of the magnetic field with the trap axis.}
We align the direction of the magnetic field with the trap axis by
maximizing the quadrupole shift as a function of the field
direction. A residual quadrupole potential $\Phi_s$ caused by
stray charges and by voltages applied to the compensation
electrodes will in general have principal axes that do not
coincide with those of the trap's quadrupole potential. This will
lead to errors in the alignment of the magnetic field $\vec{B}$
with the principal axis of the trap's electric quadrupole
potential $\Phi_{tips}$. This is due to the fact that $\vec{B}$ is
aligned with the direction $\vec{n}_{ts}$ that maximizes the
quadrupole shift of $\Phi_{tips}(\vec{n})+\Phi_s(\vec{n})$ instead
of the direction $\vec{n}_{t}$ that would be chosen without
$\Phi_{s}$. Fortunately, $\Phi_s$ is much smaller than
$\Phi_{tips}$ in the experiment. We obtain information about the
magnitude of $\Phi_s$ by (1) extrapolating the ion oscillation
frequency $\omega_z(U_{tips})$ for $U_{tips}\rightarrow 0$, (2)
measuring the electric stray field $\vec{E}_s$ and assuming the
magnitude of the field gradient to be $\propto |E_s|/r$ where $r$
is the smallest distance of the ions from the trap electrodes.
Most importantly, (3) by measuring the residual quadrupole shifts
for $\beta=0$ and another direction with $\beta=26.9^\circ$,
constraints regarding the shape, orientation and magnitude of
$\Phi_s$ can be found that allow us to calculate the average angle
$\Delta\beta$ between $\vec{n}_{ts}$ and $\vec{n}_{t}$. We find
$\Delta\beta=3^\circ$ and use this value for estimating the
uncertainty in the determination of $\Theta(3d,5/2)$.
\end{methods}


\begin{addendum}
\item[Acknowledgements] We thank H.~H{\"a}ffner for his
contributions to the optical pumping scheme and acknowledge help
with the experiment from T.~K{\"o}rber, W.~H{\"a}nsel,
D.~Chek-al-kar, M.~Mukherjee, and P.~Schmidt. We gratefully
acknowledge support by the Austrian Science Fund (FWF), by the
European Commission (SCALA, CONQUEST networks), and by the
Institut f\"ur Quanteninformation GmbH. This material is based
upon work supported in part by the U. S. Army Research Office.

\item[Competing Interests] The authors declare that they have no
competing financial interests.

\item[Correspondence] Correspondence and requests for materials
should be addressed to C.R.\newline
(email:\mbox{christian.roos@uibk.ac.at}).

\end{addendum}

\end{document}